\documentclass{llncs}
\usepackage{amsmath}
\usepackage{amsfonts}

\usepackage{xspace}
\usepackage{mathrsfs}
\usepackage{graphicx}
\usepackage{algorithmic}
\usepackage{algorithm}
\usepackage{url}
\usepackage{algorithm, algorithmic}
\floatname{algorithm}{Algorithm}
\usepackage{booktabs}
\usepackage{authblk}
\usepackage{listings}
\usepackage[usenames,dvipsnames]{color}

\author{Paolo Boldi \and Corrado Monti \and Massimo Santini \and Sebastiano
Vigna}
\institute{Dipartimento di Informatica, Universit\`a degli Studi di Milano,
Italy \email{corrado.monti@unimi.it}}

\begin{document}

\bibliographystyle{splncs03}

\title{Liquid FM: Recommending Music through Viscous Democracy\thanks{The
authors were supported by the EU-FET grant NADINE (GA 288956)}}

\maketitle

\begin{abstract}
Most modern recommendation systems use the approach of \emph{collaborative
filtering}: users that are believed to behave
alike are used to produce recommendations. In this work we describe an
application (Liquid FM) taking a completely different approach.
Liquid FM is a music recommendation system that makes the user responsible for
the recommended items. Suggestions are the result of a voting
scheme, employing the idea of \emph{viscous democracy}~\cite{viscousdem}.
Liquid FM can also be thought of as the first testbed for
this voting system. In this paper we outline the design and architecture of the
application, both from the theoretical and from the implementation viewpoints.
\end{abstract}

\section{Introduction}

Most modern recommendation systems use the approach of \emph{collaborative
filtering}~\cite{recommenderhandbook,spotify}: users that are believed to behave
alike are used to produce recommendations. The idea
behind Liquid FM is to tip over this approach by making the user responsible for
these matches: deciding who they want to resemble becomes a choice of the user,
instead of being inferred algorithmically. This scenario can be cast as a voting
scheme: each user has to select another one that is believed to be a good
recommender. This idea allows us to use this task as a testbed for
\emph{viscous democracy}~\cite{viscousdem}.

Viscous democracy is a kind of liquid democracy~\cite{o1994delegative}. In
liquid (or \emph{delegative}) democracy, each member can take an active role---by
participating directly and exercising their decision power---or a passive
role---by \emph{delegating} to other members their share of responsibility. It
can be seen as a compromise between representative democracy, where voters are
usually neglected any decision making and can only delegate others to do so, and
direct democracy, where every voter is called to an active role, regardless of
what their inclinations are.

In this sense, liquid voting systems try to take the best from both worlds.
Every member's opinion, in a direct democracy, is directly relevant to a final
decision, but the vote of each one can be (knowingly!) uninformed; instead, in a
classical representative system, elected representatives are encouraged to be
informed on the specific decision they are making, but on the other hand the
majority of people feel that their opinion on that matter is basically
irrelevant. Liquid democracy permits members to choose among expressing their opinion
directly if they feel entitled to do so, or delegating their voting power if
they believe others are more capable. Note that these two options are not
necessarily exclusive---in our case, in fact, users will be able to do both, if
they want to.

Viscous democracy was proposed by Boldi \emph{et al.}
in~\cite{viscousdem} as a particular way to compute the outcome of a liquid
democracy voting scheme. It takes advantage of known techniques for measuring
centrality in social networks, and in particular it resembles Katz's
index~\cite{katz1953new}. It stems from the assumption that the delegating
mechanism should transfer \emph{a fraction} of the user voting power. I.e., if A
delegates B and B delegates C, the trust that A puts in C should be less than if
A voted C directly. This principle will be further detailed in the next section.

This framework can be used in a variety of settings. In our application, we show
how
it can be  easily adapted to music recommendation. For a
certain music genre, we ask users to express a short list of their favorite
songs, or to delegate one of their Facebook friends they consider to be an expert on
that genre. This builds a graph of delegations for each music genre. We wish to
employ this data to create recommendations for each user.

We will detail how we extract information from this graph in
Section~\ref{sec:theory}; then, in Section~\ref{sec:practice}, we will describe
how we have developed the system: how its algorithms were implemented, the
architecture of its components, and the external resources we used; finally, in
Section~\ref{sec:conclusions} we will sum up our work and present possible
directions for future research.

\section{Viscous democracy and recommender systems}
\label{sec:theory}

From now on, we will denote with ${d_G(x,y)}$ the distance from node $x$
to node $y$ in the graph $G$, and with $o_G(x)$ the outdegree of node $x$ in $G$; we
may omit reference to $G$ if it is obvious from the context.

Let us define $U$ as the set of users and $S$ as the set of songs\footnote{As we
will explain in Section~\ref{sec:practice}, we are going to consider different
sets of songs and votes, one for each music genre treated. For the rest of this
section, we are going to consider the music genre as fixed.}. ${D=(U, A_D)}$, with
${A_D \subseteq U \times U}$, is the directed graph of delegations; an arc from
user $u$ to $u'$ means the former delegates the latter as an expert on the
topic. ${V=(U, S, A_V)}$, with ${A_V \subseteq U \times S}$, is the bipartite
graph of votes, where an arc from $u \in U$ to $s \in S$ means that the user $u$
recommends song $s$.

We are going to put some restrictions on these graphs: first of all, we are
going to assume that \label{define-friendship-graph}there is an underlying,
undirected friendship graph ${F=(U, E_F)}$, with ${E_F \subseteq U \times U }$,
where an edge ${(u, u') \in E_F}$ expresses a personal acquaintance of $u$ and
$u'$. We impose that ${A_D \subseteq E_F}$: this permits us to ensure that the
trust expressed through a delegation is a result of personal knowledge, as
suggested in~\cite{viscousdem}.

\begin{figure*}[t]
    \begin{center}
        \includegraphics[width=1.0\columnwidth]{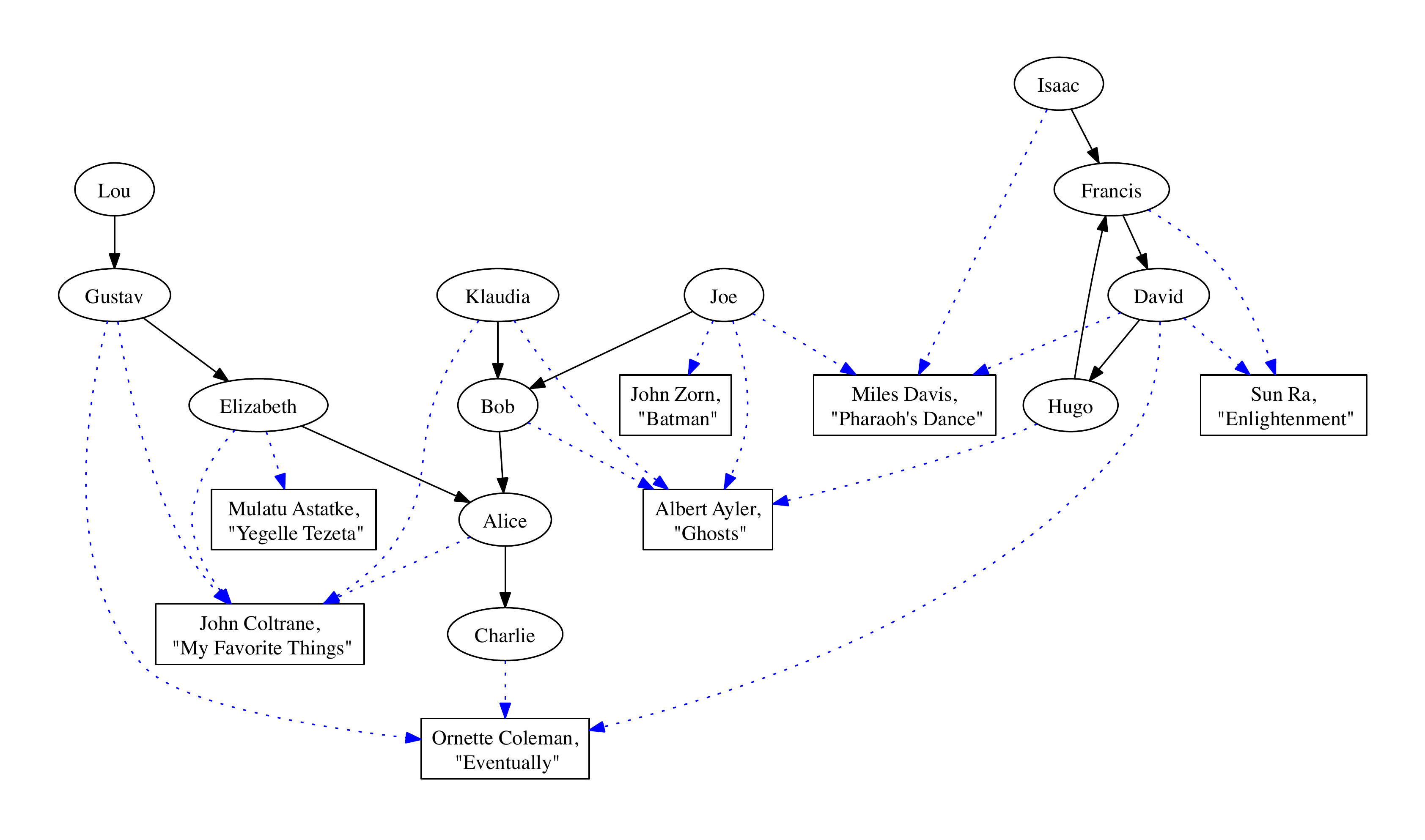}
        \caption{\label{fig:graph}
            An example of delegation and voting graphs. Users $u \in U$ are
            represented with a circle; songs $s \in S$ with a box; the
            delegation graph $D$ is drawn with black solid arrows, while the
            bipartite voting graph $V$ with blue dotted arrows.
        }
	\end{center}
\end{figure*}

Further, we are going to impose that $\forall u \in U$, we have $0 \leq o_D(u)
\leq 1$, meaning that a user can delegate only one person, and $0 \leq o_V(u)
\leq 3$, meaning that every user can vote up to 3 songs. We are going to
consider only nodes $u \in U$ having $o_D(u) > 0 \vee o_V(u) > 0$. An
example of such a setting is pictured in Figure~\ref{fig:graph}.

\subsection{Liquid voting}

A voting system is a function ${v_D: U \rightarrow \mathbb{R} }$
assigning a score to each user, depending on the delegation graph. Such a
function will be the basic building block of our recommendations.

Usually, in liquid vote this function is just the size of the tree with root in
${u \in U}$:
$$ l_D(u)= \big| \left\{ u' \in U | d_D(u',u) < \infty \right\} \big| $$
This function is used, e.g., by the well-known \texttt{LiquidFeedback}\footnote{\url{http://liquidfeedback.org/}} platform.
Nonetheless, it assumes that ``trust'' transferred from $a$ to $b$ is
the same whether $a$ delegated $b$ directly, or whether they are connected by a
long chain of delegations---and they may not even know each other.

Let us assume that we wish, instead, that the amount of trust passed on from $a$
to $b$ is greater if $(a, b) \in A_D$, and lesser if there are many steps
connecting them. To do so, we introduce a \emph{damping factor} $\alpha\ \in
(0,1]$, defining how much of the voting power of $a$ is transferred to $b$ when
$a$ delegates $b$. Therefore, the scoring function characterizing \emph{viscous}
democracy will be:
\begin{equation}
    \label{eq:v}
    v_D(u)= \sum_{u' \in U} \alpha^{d(u',u)}
\end{equation}

Authors~\cite{viscousdem} have noted how, depending on the value of $\alpha$,
the behavior of the voting function greatly differs. For higher values of
$\alpha$, the fraction of trust ``lost'' in each delegation step becomes smaller
and smaller; in fact, for $\alpha \rightarrow 1$, we have that $v_D \rightarrow
l_D$: all the nodes in the tree of $u$ contribute with all their voting power to
$u$, exactly as in pure liquid democracy. Note that if we allow $\alpha=1$, we
must explicitly avoid cycles in $D$---exactly as with pure liquid democracy;
this constraint is not needed with viscous democracy with $\alpha \in (0,1)$.

\begin{figure*}[t]
	\begin{center}
		\begin{tabular}{c|c}
			\includegraphics[width=0.45\columnwidth]{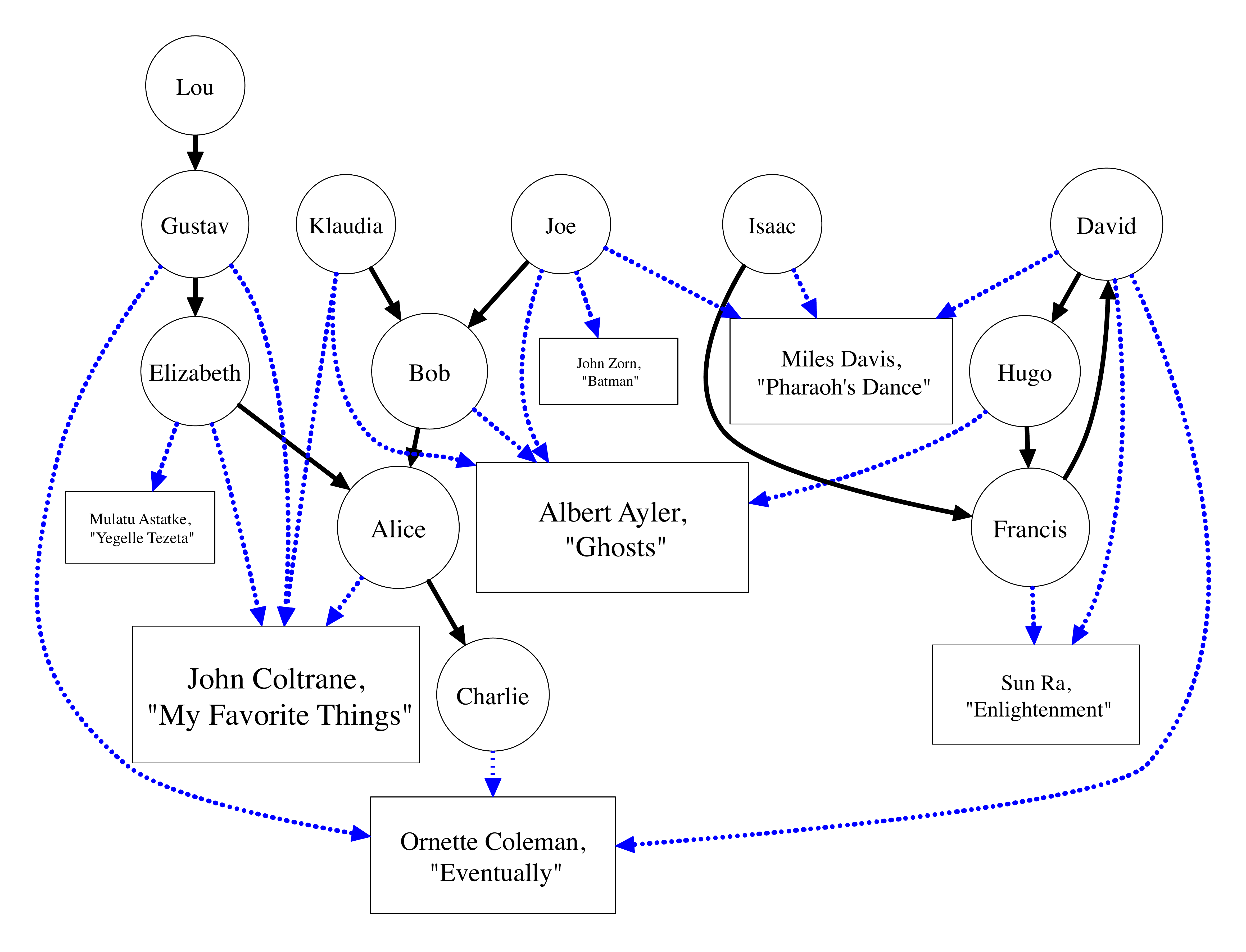}
			&
			\includegraphics[width=0.45\columnwidth]{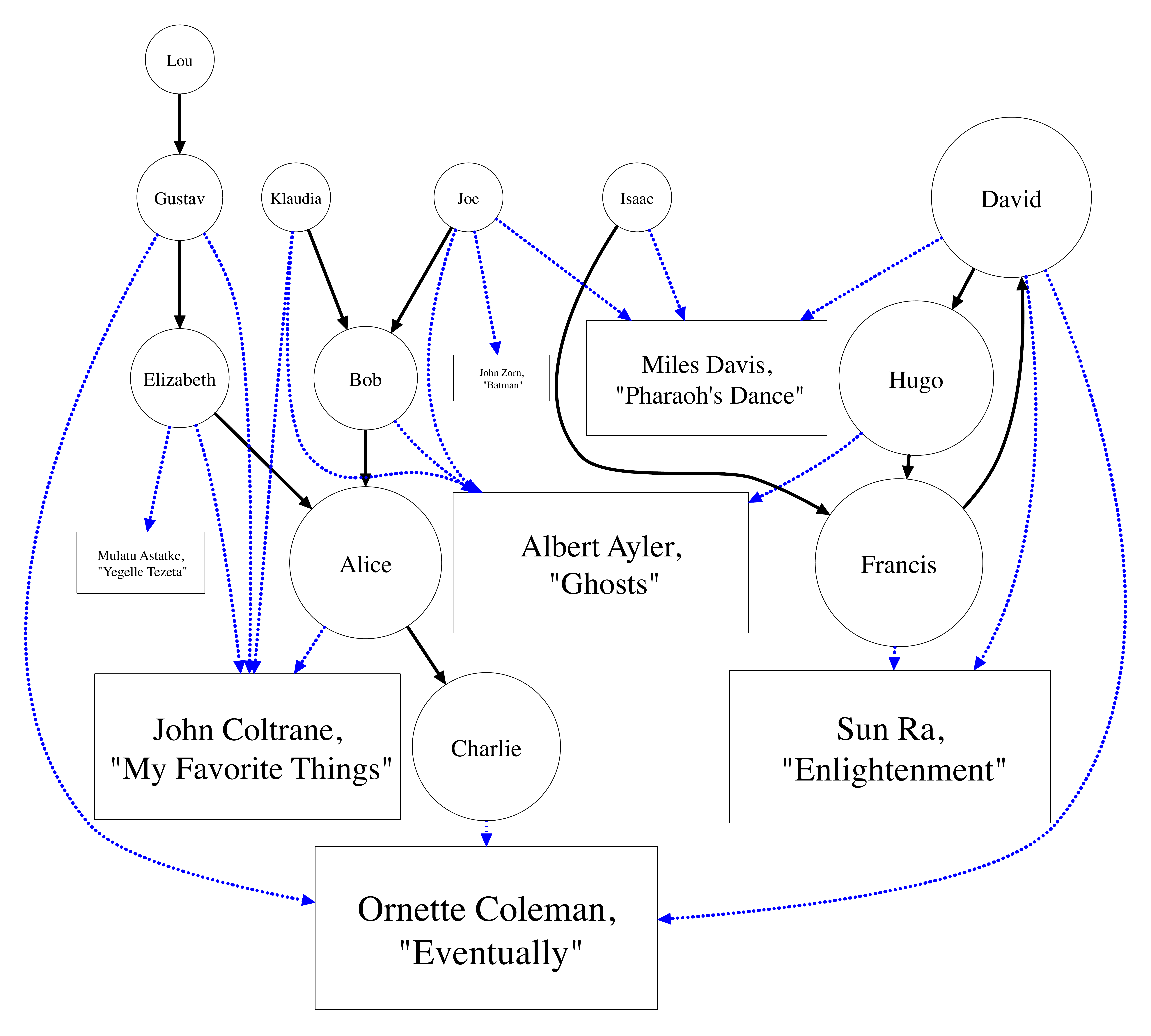}
            \\
            $\alpha=0.25$
            &
            $\alpha=0.75$
		\end{tabular}
		\caption{\label{fig:graph-scored}
            The same graphs pictured in Figure~\ref{fig:graph} are here
            displayed with node size proportional to their viscous score,
            with two different values for $\alpha$. Note how a higher $\alpha$
            gives higher importance to users delegated by important users.
            Lowering its value get us closer to a simple vote count. For example,
            Ornette Coleman's song is ranked higher than Coltrane's only for
            higher $\alpha$: this is because it is voted by fewer users, but those
            are recognized by the community as experts.
        }
    \end{center}
\end{figure*}

With $\alpha$ approaching $0$, instead, the voting power becomes
nontransferable: all users become equal, regardless of the delegations they
received; in other words, the model becomes a direct democracy, without any
proxy vote. These differences are presented graphically in
Figure~\ref{fig:graph-scored}, making use of the song-scoring function we
will show in the next section.

\subsection{Global recommendations}

Having a score for each user, we can easily score each song $s \in S$. Indeed,
we can define a function $r: S \rightarrow \mathbb{R}$ as
\begin{equation}
    \label{eq:r}
    r(s) = \sum_{u \in U | (u, s) \in A_V } v_D(u)
\end{equation}

This function will get us a score for a song proportional to the importance of
who voted it, according to $v_D$. The score is completely defined by the
graphs $V$ and $D$. We can then proceed to rank each song with $r$, and
present them to the users accordingly. As in many standard information retrieval
tasks, a user looking for results (about a certain music genre, as we
will explain in Section \ref{sec:practice}) will be presented with all possible
items---all songs in $S$---ranked from higher to lower $r$. Users will be
therefore more likely to listen to songs ranked higher in this list.

Let us call the \emph{influence} of $u \in U$ the difference the votes of user
$u$ make in the final rankings---that is, $\sum_{s \in S}{r(s) - r_{V
\backslash \{u\}}(s)}$. Please note that, since we have not normalized $r$,
users giving more votes have a larger influence in the final rankings, serving
the purpose of encouraging them to give more recommendations. However, it also
explains why we had to put a limit on $o_V(u)$: if we had not, a single user
$u$ could have an arbitrary influence on the score $r$, resulting in the
possibility of spam.

In the end, the influence of a user on song scores is determined by the number
of recommendations they give---limited, but under their control---and by the
delegations they received---unlimited, but not under their direct control.

As mentioned before, an example of how $r$ behaves is pictured in
Figure~\ref{fig:graph-scored}.

\subsection{Personalized recommendations}

The song-scoring function we presented gives the same ranks to whoever is their
observer. This behavior is unusual in recommender systems, where the goal is to
give the right recommendation to the right person. In our case, a user may be
more interested in listening to what their delegate suggested, rather than
other---possibly more popular---items. Looking at our example in
Figure~\ref{fig:graph-scored}, Francis may be more interested in listening to
``\emph{Pharoah's Dance}'', even if it is not globally highly-ranked, because it is the
recommendation of his delegate David. Similarly, Hugo may be interested in it,
because he has, in turn, delegated Francis.

This goal can be easily expressed as a personalized song-scoring function. Let us
define a function $p: S, U \rightarrow \mathbb{R}$ as
\begin{equation}
    \label{eq:p}
    p(s, u) = \sum_{u' \in U | (u', s) \in A_V} \alpha^{d(u,u')}
\end{equation}

Such a function permits the user $u$ to get a positive score only for the
songs recommended by users belonging to the chain of delegations starting in
$u$.
For the purpose of maintaining this intention, but at the same time avoiding to
completely discard all the songs highly ranked by the original $r$, we can
define a linear combination of the two functions, normalized to $1$:
\begin{equation}
    \label{eq:c}
    c(s, u) = \delta \frac{p(s, u)}{\max\limits_{s' \in S} p(s',u) } +
    (1 - \delta) \frac{r(s)}{\max\limits_{s' \in S} r(s') }
\end{equation}
where $\delta \in [0, 1]$ regulates the amount of personalization of $c$.

\begin{figure*}
	\begin{center}
		\begin{tabular}{c|c}
			\includegraphics[width=0.45\columnwidth]{figures/graph-75.pdf}
			&
			\includegraphics[width=0.45\columnwidth]{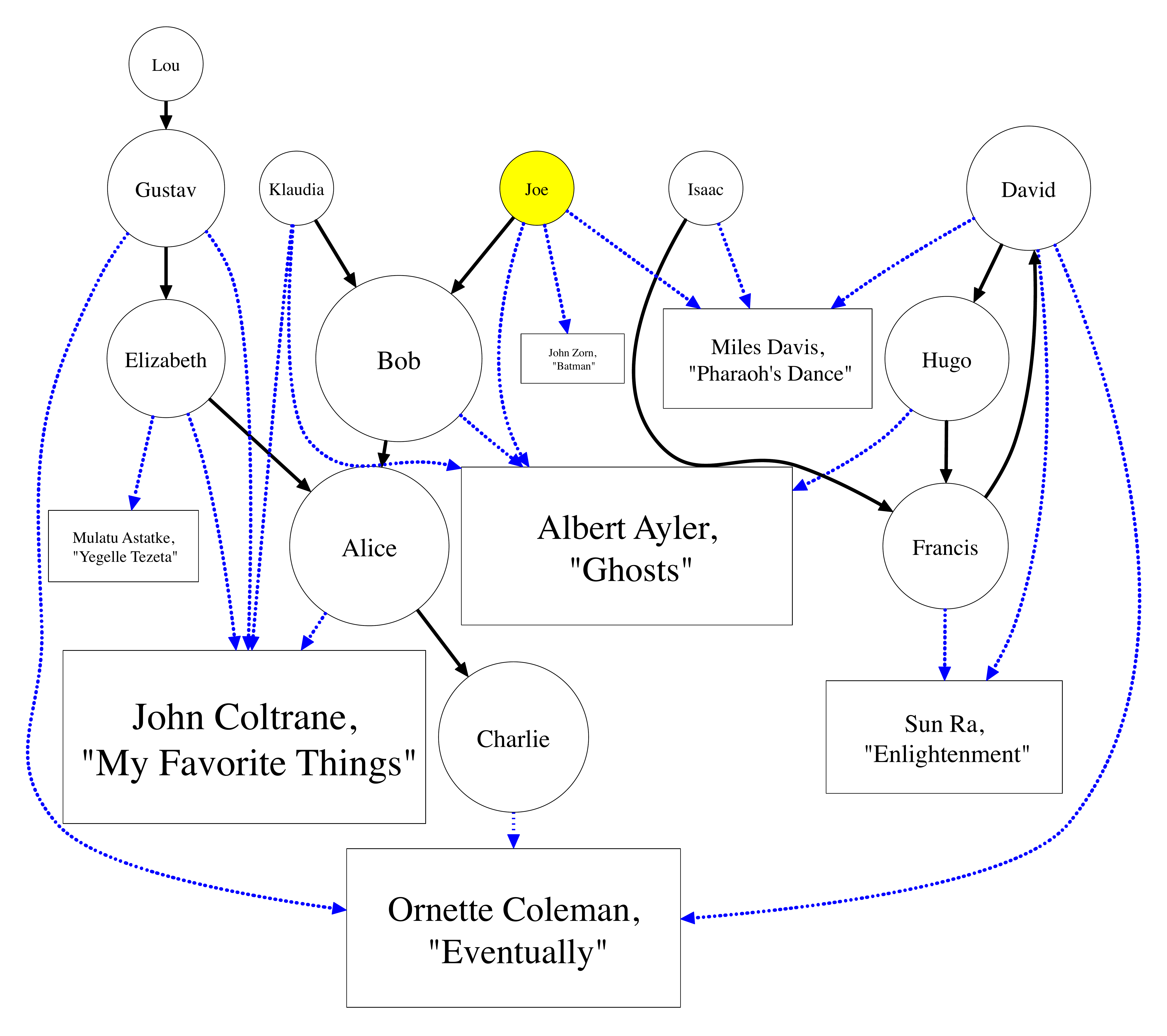}
            \\
            $r(\cdot)$
            &
            $c(\cdot, \mathtt{Joe})$
		\end{tabular}
		\caption{\label{fig:graph-personal}
            The same graphs pictured in Figure~\ref{fig:graph-scored} are here
            displayed with global song-scoring function $r$ on the left and, on
            the right, with the personalized function $c$ from the view point of
            user Joe (in yellow) and $\delta=0.9$. In the latter,
            recommendations suggested by the delegate of Joe acquire more
            importance; those suggested by indirect delegates (namely, Alice and
            Charlie) increase as well, but by a minor amount.
        }
    \end{center}
\end{figure*}

An example is pictured in Figure~\ref{fig:graph-personal}.

\subsection{Insights for users}

In addition to the presented ways to compute recommendations, the setting here
described also permits to compute other information that may be of interest to
the users. Particularly, it allows them to know how authoritative (i.e.,
trustable) their taste is in a particular music genre. The function $v_D$, in
fact, can be normalized into a percentile-based scoring, obtaining an easy-to-read assessment
in the form ``$u$ is better than $\widehat{v_D}(u)$ people out of 100'' (for a
specific genre), with $\widehat{v_D}(u)=100 \frac{|\{u' \in U | v_D(u') <
v_D(u)\}|}{|U|}$. It can then be used to provide useful information from two
different perspectives:

\begin{enumerate}
    \item Showing to the user a fair evaluation about which music genres they
    are believed to be more expert about.

    \item Presenting to a user interested in learning more about a specific
    genre which of their friends is considered an expert---making use of
    the direct knowledge graph defined on
    page~\pageref{define-friendship-graph}.
\end{enumerate}

\section{Development}
\label{sec:practice}

We will now discuss how the presented techniques have been implemented in
practice. The final result is Liquid FM: a Facebook application that enable its
users to vote one of their friends as an expert on a music genre, and (by means
of the described formulas) recommends them some piece of music to listen to, by
identifying the best experts.

Firstly, we will present a general overview of the architecture of Liquid FM,
explaining the role of its main components; then, we will give a more detailed
look at the implementations of the formulas presented in
Section~\ref{sec:theory}; finally, we will discuss the external components we
employed.

\paragraph{Categories}

As anticipated in the previous section, we applied our scoring algorithms to
9 music genres, called \emph{categories} from now on, and their set will be
denoted as $C$. In this way, we will have different votes and different
recommendations for each category. Such a behavior is closer to reality: an
expert in HipHop is not assumed to be qualified to give, say, classical music
suggestions. However, it also permits to have different graphs for the same
users---an interesting fact for future analysis.

The selected categories were Classical, Electronic, Folk, HipHop, Indie, Jazz,
Metal, Pop, Rock. They were chosen by inspecting LastFM top 20
tags\footnote{\url{http://www.last.fm/charts/toptags}} and discarding those not
expressing a musical genre (such as ``seen live'') and sub-genres (having
included Indie and Rock, we discarded ``Indie Rock''). We decided to add
Classical (only ranked 36th on Last Fm), since it is a different and interesting
community, under-represented in services such as the one we referred to.

\subsection{Overview}

\begin{figure}[t]
    \begin{center}
        \includegraphics[scale=0.2]{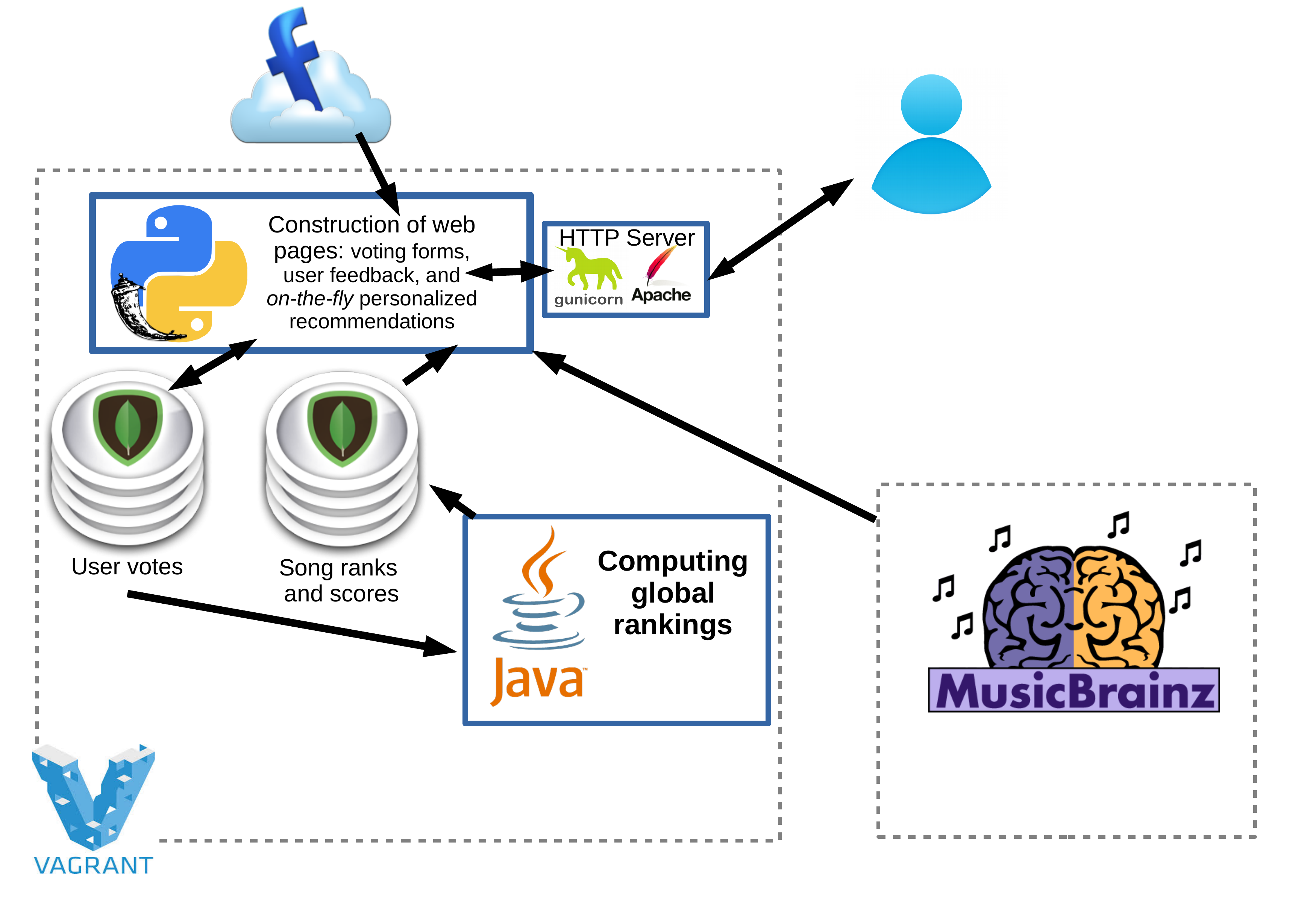}
		\caption{\label{fig:architecture}
            A schematic representation of the architecture of Liquid FM. An arrow
            going from A to B indicates data flowing from A to B.
        }
	\end{center}
\end{figure}

As pictured in Figure~\ref{fig:architecture}, Liquid FM features two main
components:

\begin{itemize}
    \item a Java part, with the role of analyzing the whole graphs and computing
    global scores through $v_D$ and $r$ (equations~\ref{eq:v} and
    ~\ref{eq:r}): it is meant to be fast, and executed periodically;

    \item a Python part, with the role of glueing the different parts together
    and providing all the other functions: from the construction of web pages to
    the implementation of personal scores (functions $p$ and $c$,
    equations~\ref{eq:p} and~\ref{eq:c}).
\end{itemize}

These two parts interact with each other through a shared database,
that persistently stores every information. We chose MongoDB, an open-source
document-oriented NoSQL database, for various reasons:

\begin{itemize}

    \item We want fast access in reading and writing data (especially very
    small chunks, as in delegations and votes) in order to be able to
    support a large amount of users, like in modern recommendation systems.
    Moreover, we would like our system to be scalable.

    \item We want flexibility: since this application is also a proof-of-concept,
    we need to be able to modify data schemas, totally or partially, without
    much concern.

    \item Finally, we do not often need complex operations, involving more than
    one collection. When it occurs, we would like to control what is happening
    at application-level, permitting fine-grained handling.

\end{itemize}

On this database, we have two main collections gathering user-submitted data:
following the notation introduced in Section~\ref{sec:theory}, the first
stores the graph $D$ and the second the graph $V$. These collections are both
represented in Figure~\ref{fig:architecture} as ``User votes''. A document in
the collection for $D$ looks like this\footnote{Whenever the id of a document is
not explicitly expressed, it is automatically generated by MongoDB. This is done
efficiently; furthermore, such an id stores the timestamp of creation of the
document.}:
\lstset{escapeinside={[*}{*]}}
\begin{lstlisting}
{   category : [*$c$*], from : [*$u$*], to : [*$u'$*]     }
\end{lstlisting}

While a document in the collection for $V$ has this structure:
\lstset{escapeinside={[*}{*]}}
\begin{lstlisting}
{   category : [*$c$*], user : [*$u$*], advice : [*$s$*]    }
\end{lstlisting}

\begin{figure*}
    \begin{center}
        \includegraphics[scale=0.2]{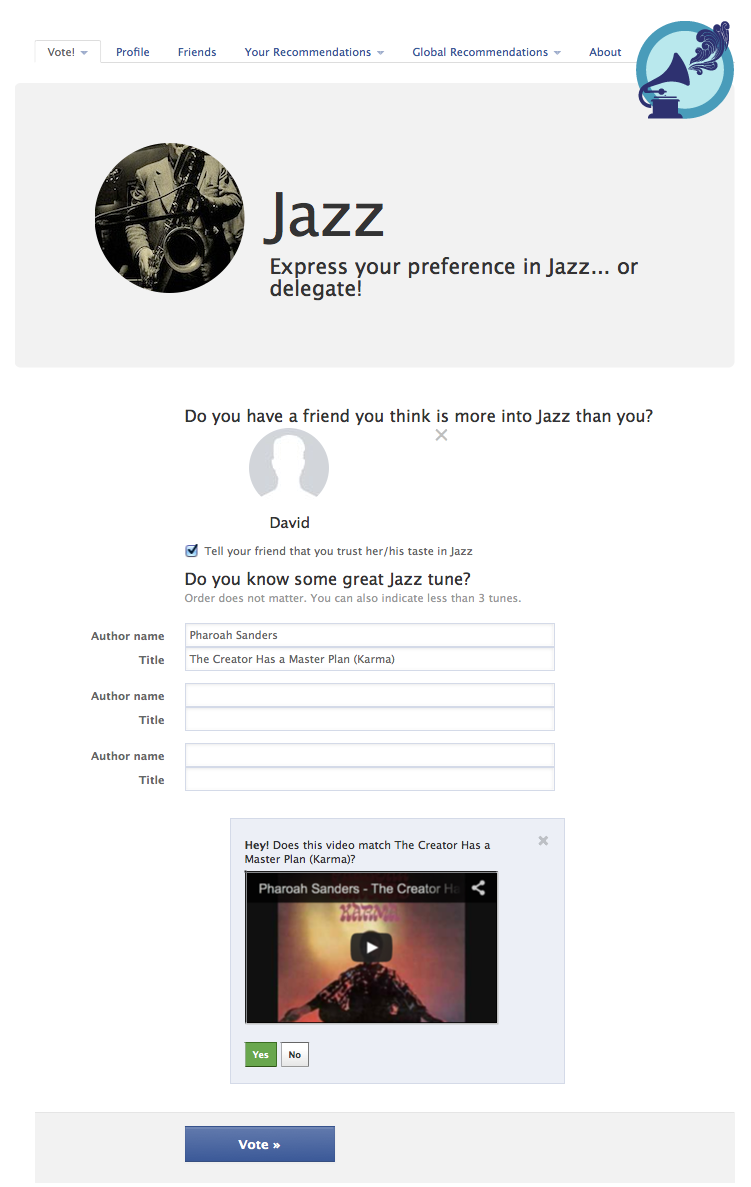}
		\caption{\label{fig:screenshot-voting}
            A screenshot of the voting phase of a user.
        }
	\end{center}
\end{figure*}

The advice $s$ is a dictionary containing author and title of the song, as
well as a YouTube video id. In fact, we associate with each song selected by a
user a YouTube video, in order to be able to play it as a recommendation.
YouTube is in fact one of the largest and most used music streaming platforms,
and it can be included in third-party services (with small limitations). A
screenshot of the voting phase in displayed in Figure~\ref{fig:screenshot-voting}.

Please note that the structure of an advice, as well as the category $c$, is
well-incapsulated: therefore, the schemas of these collections can be easily
extended in the future in order to support different (i.e., not music-related)
scopes.

\subsection{Recommendations}
The division of global and personalized recommendations into two separate
components originates from efficiency reasons. Having to compute and store all
the personalized scores for each user would be impracticable, as they would
be $|C| \cdot |S| \cdot |U|$ scores. Therefore, they are computed with a lazy
approach: when a user $u$ asks for her personal recommendations, we compute
all of them \emph{on-the-fly} and cache them. Global recommendations, instead,
are the main result of the system, and every user depends on them---even to see
the personalized scores, since we use the function $c$ (equation~\ref{eq:c}).
For this reason, we compute them periodically with a fast Java component, and save
them to a dedicated MongoDB collection.

\paragraph{Global recommendations in Java}
The global recommendation component was carried out in Java, since for this
task it is faster than Python and since efficient
open-source libraries to deal with graphs are available; in
particular, we employed extensively the \mbox{WebGraph}
framework~\cite{webgraph-i} and the~\texttt{fastutil} library.

This component is run periodically. It takes as input the graphs $D$ and $V$,
memorized in their MongoDB collections, and it results in a new collection for
each category, composed of documents of this form:
\lstset{escapeinside={[*}{*]}}
\begin{lstlisting}
{   advice : [*$s$*], rank : [*$r(s)$*] }
\end{lstlisting}

and in another collection ranking users, where each document has this form:
\begin{lstlisting}
{   _id: [*$u$*],
    [*\emph{\textit{category}}$_1$*]: { score : [*$v_D(u)$*], perc : [*$\widehat{v_D}(u)$*] }, [*$\dots$*]
}
\end{lstlisting}

We can schematize the process, for each category $c$, in these steps:
\begin{enumerate}
    \item Read the graph $D$ from MongoDB and convert it to WebGraph format.

    \item Use a parallel implementation of the Gauss-Seidel method (from
    WebGraph) to compute $v_D$ for each user $u$. We decided to choose 0.75 as
    the value of $\alpha$.

    \item Compute the percentile-based normalization $\widehat{v_D}$, and save
    the user-ranking collection to MongoDB.

    \item Read the graph $V$ from MongoDB, identifying the set $S$ of
    songs. In this step, we also find which YouTube video is the most
    frequently associated with a certain song, using author and title as
    identifiers.
    While doing this, we compute $r(s)$ for each song $s$.

    \item Save $r(s)$ in their collection, indexing documents by decreasing
    scores. Also save $\max\limits_{s \in S} r(s)$, for normalization purposes.
\end{enumerate}

\paragraph{Personalized on-the-fly recommendations}
As discussed above, personalized recommendations are computed on-the-fly by a
Python component. Python was in fact chosen as the main language of the
application, due to its versatility and its fast production times; also,
we decided to use \texttt{Flask}\footnote{\url{http://flask.pocoo.org/}}, an open-source web development micro-framework
particularly suited for our task.

Personalized recommendations are computed only when users ask for them, since
they require to see only a very small part of the graphs, and because storing
all of them would be unfeasible. The score we will use to rank personalized
recommendations is the function $c(s, u)$ (eq.~\ref{eq:c}); in order to compute
it we must, in the first place, compute $p$ (eq.~\ref{eq:p}).

To compute $p(s,u)$ for all songs $s \in S$ and a fixed user $u$ we walk through
the chain of delegations on graph $D$, starting from $u$. Since $\forall u$
$o_D(u) \leq 1$, this path on $D$ is unique (although it may end in
a cycle). Therefore, we simply proceed as follows (for a suitable stopping
threshold $\epsilon$):

\begin{algorithm}
    \begin{enumerate}
        \item Let $\texttt{p}$ be a map with 0 as default value
              for missing keys.
        \item \textbf{while}
                $ t > \epsilon$ \textbf{and}
                $\exists u'$ s.t. $(u, u') \in A_D$ \begin{enumerate}
            \item $u \leftarrow u'$ s.t. $(u, u') \in A_D$ \item For each
            $s$ s.t. $(u, s) \in A_V$: \\
                    \-\ $\texttt{p} [ s ] \leftarrow \texttt{p} [ s ] + t$
            \item $t \leftarrow t \cdot \alpha$
        \end{enumerate}
    \end{enumerate}
\end{algorithm}

Now we have all the ingredients for function $c$, and we can proceed to compute
the ranking order according to it.

First of all, consider that the ranking order induced by $c(s)$ is equivalent to
\[
\bar{c}(s) = k \cdot p(s, u) + r(s)
\qquad\text{where}\qquad
k = \frac { \delta       \cdot \max\limits_{s' \in S} r(s')    }
          { (1 - \delta) \cdot \max\limits_{s' \in S} p(s', u) }
\]

Therefore, for each element $s$ of the map $\texttt{p}$,
we multiply its value by $k$ and add the value of $r(s)$.
Then, we retrieve all the other
items $s$ s.t. $r(s) \geq \min\limits_{s'} \texttt{p}[s'] $,
and insert them in the map $\texttt{p}$.
Finally, we can build the iterator of personal recommendations by chaining two
iterators:

\begin{enumerate}
    \item the iterator of all elements in $\texttt{p}$, sorted by their values;
    \item the iterator of all other elements $s \in S$ having
    $r(s) < \min\limits_{s'} \texttt{p}[s']$, sorted by their values; remember
    that they are already indexed in this order in the database.
\end{enumerate}

The final iterator can be implemented in a lazy fashion, allowing us to retrieve
the elements of the second iterator only when necessary. The first iterator,
instead, will be computed eagerly upon its request, and then cached. To cache
these (and other) values, we employed \texttt{redis}\footnote{\url{http://redis.io/}}, an open-source in-memory
key-value cache.

\subsection{External services}
To conclude this section, we would like to briefly describe the main external
software components we used in developing Liquid FM.

\paragraph{Facebook}
As mentioned earlier, Liquid FM is a Facebook application. The reason for it is
that we used the Facebook friendship graph as the graph $F$ defined on
page~\pageref{define-friendship-graph}. In fact, Facebook is at the moment the
largest existing social network (with 1.4 billion users), and it has been
previously used as a good approximation of an acquaintance
graph~\cite{fourdegrees}. Therefore, we require users to have a Facebook
account, in order to limit their choice of delegate to their acquaintances. Accordingly,
in the collections described earlier, we used a Facebook-provided id\footnote{To
protect users' privacy, this id is valid only within our app, and cannot be used
outside of it.} to identify a user $u$.

\paragraph{Musicbrainz}
To ensure the validity of the set $S$ of songs chosen by users, we check them
against the Musicbrainz database. Musicbrainz is an open music database that
anyone can edit. At the time of writing, it contained information about
more than $900\,000$ artists and $14\,000\,000$ recorded songs. Since it follows
the open-content paradigm, a user who does not find its favorite song in the
database is in principle free to add it; however, the
community-review process acts as a filter.
Furthermore, Musicbrainz provides a disk image to set up a
virtual machine with a fully-functioning Musicbrainz server; we used this
approach to be able to access the database fast, without network delays and
minimizing the impact on their hosts. Moreover, the database of this virtual
machine has been set up to self-update itself periodically, in order to adopt
every new edit accepted by Musicbrainz.

\section{Discussion, conclusions and future work}
\label{sec:conclusions}

In this work, we presented a Facebook application aimed at putting
the viscous democracy framework~\cite{viscousdem} to the test. This is at the
same time a proof-of-concept of how that voting system can be practically
implemented in a real-world social network, and a way to collect data
corroborating (or disproving) the supposed advantages of viscous democracy when
compared to other, more standard, ways of performing elections in a social
setting.
An interesting point, here, is that the usage of viscous democracy for
recommendation seems to avoid the filter bubble~\cite{pariser2011filter}, at
least in its more algorithmic sense, because this kind of recommendation does
not rely on collaborative filtering but is based on a conscious choice. Whether
this choice (delegation) can itself induce a similar kind of bubble will be
subject of future analysis.

The discussed application is currently active on \url{http://bit.ly/liquidfm},
and we have so far collected some small datasets; currently, the delegation graphs
consist of few tens of delegations, so it is impossible to draw any conclusion
from them. In order to be able to collect larger amount of information it is
crucial that
we find a way to make the application \emph{viral}: this is a matter of
social engineering that needs to be taken into careful consideration.

\bibliography{biblio}
\end{document}